\documentclass{article}
\usepackage{spconf,amsmath,graphicx, tabularx}

\title{Adapting Large Language Model with Speech for Fully Formatted End-to-End Speech Recognition}
\name{BLIND}
\address{BLIND}
\name{Shaoshi Ling$^{\dag}$, Yuxuan Hu$^{\dag}$, Shuangbei Qian$^{\dag}$, Guoli Ye$^{\dag}$, Yao Qian$^{\dag}$, \textit{Yifan Gong, Ed Lin, Michael Zeng}\thanks{$^{\dag}$Equal contribution}}
\address{Microsoft Could and AI}
\begin{document}
%\ninept
%
\maketitle
\begin{abstract}
Most end-to-end (E2E) speech recognition models are composed of encoder and decoder blocks that perform acoustic and language modeling functions. Pretrained large language models (LLMs) have the potential to improve the performance of E2E ASR. However, integrating a pretrained language model into an E2E speech recognition model has shown limited benefits due to the mismatches between text-based LLMs and those used in E2E ASR. In this paper, we explore an alternative approach by adapting a pretrained LLMs to speech. Our experiments on fully-formatted E2E ASR transcription tasks across various domains demonstrate that our approach can effectively leverage the strengths of pretrained LLMs to produce more readable ASR transcriptions. Our model, which is based on the pretrained large language models with either an encoder-decoder or decoder-only structure, surpasses strong ASR models such as Whisper\footnote{https://github.com/openai/whisper}, in terms of recognition error rate, considering formats like punctuation and  capitalization as well. 

\end{abstract}
\section{Introduction}
\label{sec:intro}

A fully-formatted End-to-End ASR transcription model converts speech signals into complete transcriptions that include punctuation, capitalization, numbers, and other formatting information. This makes the transcription more readable and aligns with the expectations of the users. It is also preferred for certain downstream NLP tasks such as machine translation (MT), summarization, and natural language understanding (NLU), where models are typically trained using large amounts of written text data. Conventional ASR systems rely on post-processing to restore these formats. However, post-processing modules are typically based on text-to-text conversion alone. Acoustic clues such as prosody, which can help predict the use of question marks or other formatting, are not well utilized. Consequently, this could lead to erroneous format conversion, even when done manually.

%This can result in inaccurate format conversion, even when done manually.

E2E modeling for speech-to-full-transcription requires large amounts of high-quality paired training data. SPGISpeech \cite{o2021spgispeech}, a publicly available dataset containing 5,000 hours of transcribed financial audio, is a good choice for research on fully-formatted E2E ASR transcription. The data quality is high, with fully-formatted professional manual transcriptions and diverse coverage of speakers with both L1 and L2 accents. However, 5,000 hours of speech in a single finance domain is not sufficient for modern research on multi-domain, generalizable ASR modeling. Whisper \cite{radford2022robust} was trained in a supervised manner on 680,000 hours of audio and the corresponding transcripts collected from the internet. Unlike conventional ASR modeling, no text normalization was performed on transcriptions, so Whisper can predict text in its transcribed form. Although Whisper is a large-scale speech recognition model in terms of training data size and model size, its scale remains relatively small compared to modern text-based LLMs.

In this paper, we propose to adapt a pretrained LLM with speech for fully formatted E2E ASR transcription tasks. The main contributions of our work are:
\begin{enumerate}
    \item We employ a composable model architecture, in which speech encoder can be integrated into a pre-trained LLM with either an encoder-decoder or a decoder-only structure. This design allows us to fully leverage the strengths of both the speech and language models with minimal changes to the architecture.
    %\item We propose a CTC-based down-sample method to address the length mismatch between speech and text representations.
    \item Instead of using an up-sampling method for text representations like in previous works, we propose a CTC-based down-sampling method for speech representation to address the length mismatch between speech and text representations. Our approach endeavours to align the representation spaces of both speech and language more closely in an end-to-end manner.
    \item LLMs are typically trained using written text. We leverage their output directly to avoid the need for format post-processing in conventional ASR. Our results demonstrate the benefits of using a pre-trained LLM for fully-formatted ASR transcription tasks across various domains.
    \item To the best of our knowledge, we made the first attempt to enhance fully formatted ASR transcription with a pretrained LLM in an E2E manner.
    %\item We demonstrate the gains of using pre-trained LMs for the fully-formatted ASR transcription tasks across various domains.
    %\item We conduct a sophisticated analysis of the error patterns in fully-formatted transcriptions produced by SOTA ASR systems.
\end{enumerate}

\section{Related Work}
E2E speech recognition models were typically trained by using paired speech and text. However, unpaired text-only is much more available than the paired data. Many studies have explored ways to utilize the text data to improve the performance of E2E ASR. A straightforward approach is to leverage an external language model (LM) or even a LLM in the beam-search-based inference stage through LM fusion methods such as shallow fusion \cite{chorowski2016towards, hu2023massively}, cold fusion \cite{sriram2017cold,toshniwal2018comparison}, and deep fusion \cite{gulcehre2015using}. Another approach is to inject text data into pretraining to learn unified speech-text representations through multi-task learning, including tasks such as speech-to-text recognition, speech-to-text translation, text-to-text translation, masked language modeling, etc \cite{tang2021general,tang2022unified,slam,maestro,usm}. To address the difference in length between speech and text representations when learning a unified speech-text representation, text representations can be up-sampled according to the corresponding phoneme duration in the paired speech \cite{tang2021general, maestro}. Alternatively, speech representations can be down-sampled to match the length of text representations using a length adaptor \cite{li2020multilingual}. Meanwhile, a pretrained LM or LLM have also been employed as seed models or used to initialize part of the weights. To facilitate training, the speech was represented by using discrete tokens that were combined with text tokens to form a finite vocabulary \cite{rubenstein2023audiopalm}. While it has significantly improved performance on tasks such as speech-to-text translation, the gains on speech recognition have been limited and, in some cases, have even resulted in degradation by comparing with conventional E2E ASR models.    

The development of conventional ASR systems involves a series of data pre-processing steps, such as text normalization (TN) for ASR modeling, and post-processing steps, such as inverse text normalization (ITN), to produce the final outputs. The post-processing steps largely depend on the speech domains and the customer’s specifications \cite{gandhi2022esb}.  E2E fully formatted speech recognition was first introduced in a study \cite{o2021spgispeech} where autoregressive conformer-transformer and non-autoregressive conformer-CTC models were used to learn the mappings from acoustic features to token sequences of complete English text appeared in U.S. print publications. This full ASR transcription requires the post-processing steps including ITN, punctuation, capitalization and disfluency removal in the conventional ASR systems.  The authors demonstrated that their proposed approach is feasible and both models can produce a decent performance on their offered SPGISpeech corpus. The transformer-based encoder-decoder model, Whisper \cite{radford2022robust}, with the off-the-shelf transformer-transformer architecture  has also been shown to outperform conventional systems when large amounts of supervised data are available. There have been a few works \cite{liao2023improving, choe2019neural} to leverage a pretrained LM to improve the readability or grammatical error correction for ASR transcriptions, but all have been based on post-processing in a second-pass stage. %To the best of our knowledge, our work is the first attempt to enhance fully formatted ASR transcription with a pretrained LLM in an E2E manner.

\section{Method}
Our model architecture is composable, allowing for a speech encoder to be plugged into a pre-trained LLM that has either an encoder-decoder or a decoder-only structure.
\subsection{Encoder-decoder based LLM}
\begin{figure}
\centering
\includegraphics[width=0.35\textwidth]{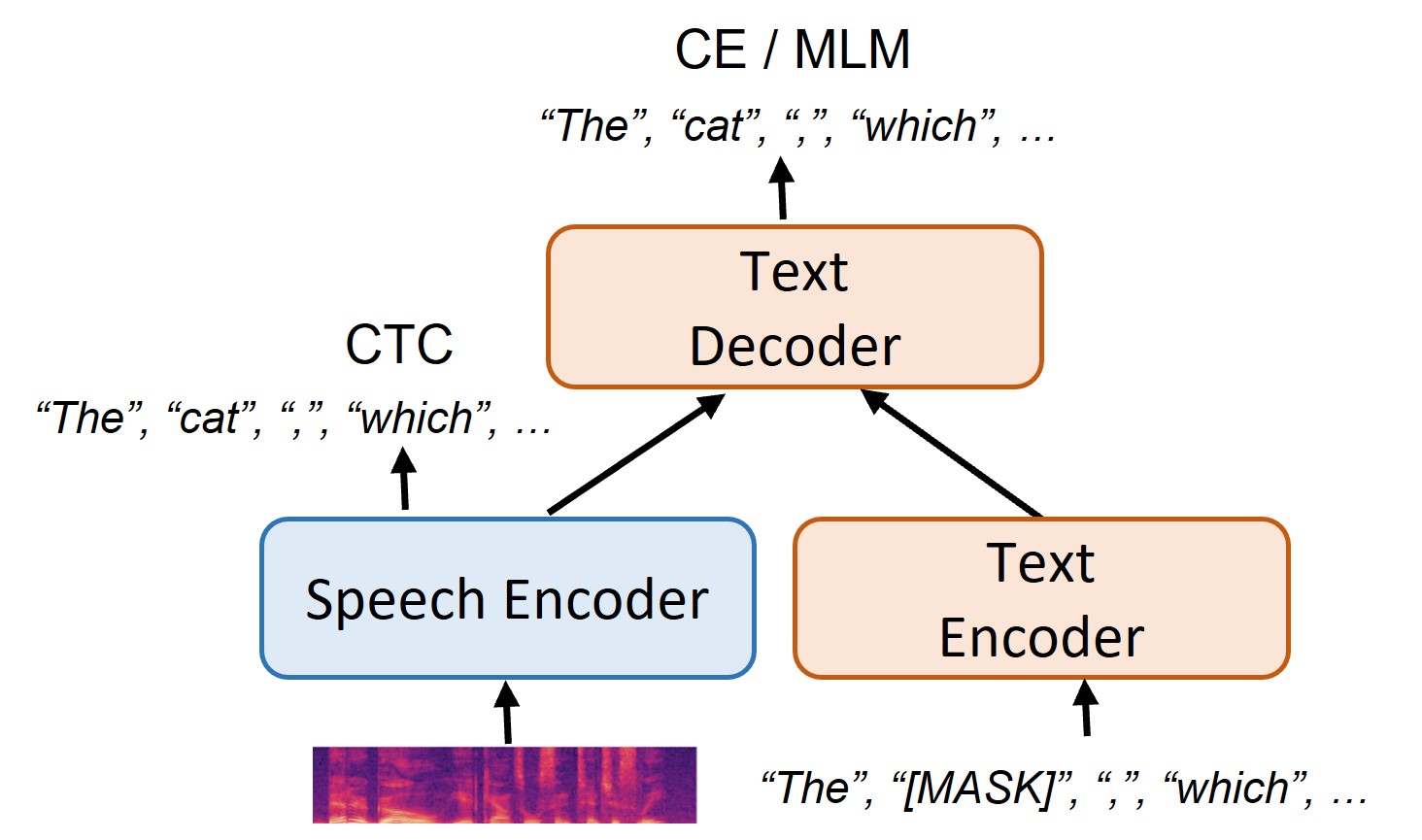}
\caption{The composition of speech encoder and encoder-decoder based LLM}
\label{fig:encoder-decoder}
\end{figure}

The composition of speech encoder and encoder-decoder based LLM is illustrated in Figure \ref{fig:encoder-decoder}, where the representations output from the speech encoder are fed into the text decoder.   

The text encoder and text decoder are initialized using a pretrained LLM. The text tokenizer of this pretrained LLM is also utilized for the speech recognition task. Given that the token size of the LLM is usually much larger than that of speech recognition, the token embedding layers for both input and output tokens are fixed to avoid insufficient training. The text encoder serves its purpose only during training to help train a better decoder by leveraging text-only data and it is not required during the inference stage.

%and the text tokenizer of the pretrained LLM is also applied to the speech recognition task (CE loss). 

During the training phase, three types of loss: Connectionist Temporal Classification (CTC), Cross-Entropy (CE), and Masked Language Modeling (MLM), are utilized to learn fully-formatted transcription prediction from both paired speech-text data and text-only data.  CTC loss is implemented on the output of the speech encoder via an additional linear projection into the target token space. Meanwhile, the MLM and CE losses are applied to the outputs of the decoder. Each batch is composed of two mini-batches: one containing paired speech and text samples, trained with CTC and CE losses,  while the other contains text-only samples, trained using the MLM loss. Gradients are accumulated within a single batch and, once two mini-batches are processed, the accumulated gradients are back-propagated and the model parameters are updated.

\subsection{Decoder-only based LM}

\begin{figure}
\centering
\includegraphics[width=0.3\textwidth]{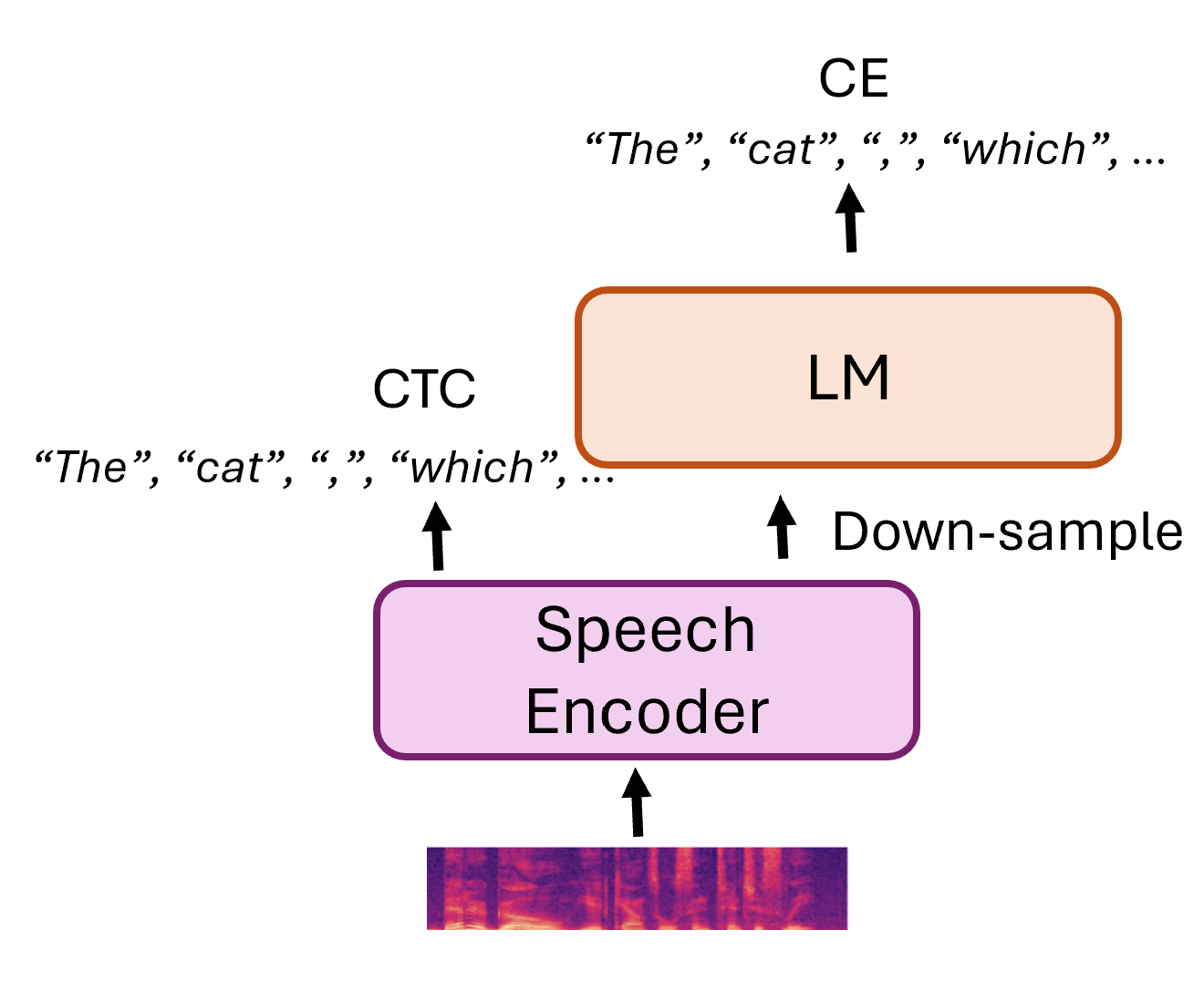}
\caption{The composition of decoder-only based LLM}
\label{fig:decoder}
\end{figure}

As decoder-only based LLM like GPT-series \cite{radford2019language, brown2020language}, LLaMA \cite{touvron2023llama} became dominated recently, we also explore its use for speech recognition task. The key components are shown in Figure \ref{fig:decoder}. To effectively leverage the pre-trained language model, we've made minimal architectural changes to it. We add LoRA \cite{hu2021lora} adapter to the pre-trained LLM. The LoRA adapter is designed to reduces the number of trainable parameters by learning pairs of rank-decompostion matrices while keeping the original weights unchanged. We then proceed to and train all other parameters in the end-to-end manner. 

%In our scenario, we provide a sequence of hidden states from the speech encoder as the prompt. We anticipate that the pre-trained language model will produce the desired transcript.

Our intuition is derived from the error-correction task where given a sequence of tokens, the LLM can produce the correct form with a appropriate prompt. In our scenario, we provide a sequence of hidden states from the speech encoder as the prompt. We anticipate that the pre-trained LLM will produce the desired transcript. However, the speech sequence often differ significantly 
% than
from 
text sequence the LLM has encountered during its training phase. There are two notable distinctions. The first is a mismatch in length. As we know, the number of frames in a speech sequence can greatly exceed the length of its transcript. To minimize this length discrepancy, we down-sample the hidden states from speech encoder by using CTC blank probability \cite{tian2021fsr, wang2023accelerating}. Specifically, we discard any frames with a CTC blank probability exceeding a set threshold. By employing a substantial blank threshold, we can effectively align the length of the transcript and the speech hidden states. The second distinction lies in the information contained within the speech hidden states and text embeddings. To effectively bridge the speech and text modalities, we replace the CTC layer with the input embedding layer from pre-trained LLM. This strategy could encourage the hidden states from speech encoder to align more closely with the text input embeddings from the LLM of the corresponding transcript.

\section{Experiments}
\subsection{Data and Metrics}
%75,000 hours of audio with human captions were collected for this study. It covers a diverse range of domains such as education, entertainment, sports, medicine, health, etc. 
%This study gathered 75,000 hours of audio with fully formatted corresponding human captions, covering a diverse range of domains, including education, entertainment, sports, medicine, health, and more. We followed the data processing methods introduced in \cite{radford2022robust}. 
This study  used 75,000 hours of transcribed speech, covering a diverse range of domains, including education, entertainment, sports, medicine, health, and more.
We retained the raw text of the transcription without any preprocessing typically used in conversational ASR modeling. Audio files were chopped into 30-second segments and paired speech-text training data were formed using segments and their corresponding segments.

The dialogistic dataset\footnote{https://huggingface.co/datasets/esb/diagnostic-dataset} of the ESB benchmark \cite{gandhi2022esb} used to assess the performance of E2E ASR systems, was employed as a testing set to evaluate the performance of our models. The transcriptions of this dialogistic dataset follow a strict orthographic and verbatim style guide. To conform to the conventions of written text and the transcriptions in SPGISpeech, we manually removed disfluencies such as partial words, repeated words, and interjections. The dialogistic dataset consists of a total of 8 datasets. We removed the AMI dataset due to many empty transcriptions after removing disfluencies. Additionally, we replaced SPGISpeech data with 1,000 utterances from the original SPGISpeech development set. In the original SPGISpeech transcription, the disfluency was removed. However, it was added back when used in the dialogistic dataset. So we used the original SPGISpeech to avoid manual work.  Both SPGISpeech and Earning22 were derived from company earnings calls, so we also removed Earning22 from the evaluation sets to avoid duplication of domains. The data sets used for our evaluation, along with their corresponding numbers of utterances and tokens, are listed in the table \ref{TER}. We plan to release this evaluation set for public use. 

None of the data from these evaluation data sets were used during model training, which represents a zero-shot setting for evaluation. We assess the generalization capabilities of our models as well as public Whisper models \footnote{https://huggingface.co/openai/whisper-large} on this evaluation set. The performance of these models are tested by using Token Error Rate (TER) between the predictions and the ground-truth transcriptions. Here a token is defined as a unit separated by spaces in a sentence. Similar to the orthographic WER used in \cite{o2021spgispeech,gandhi2022esb}, TER is both punctuation and case-sensitive. 
%Any discrepancies in punctuation and case, such as omissions, inaccuracies, or superfluous additions in the constituent words, are counted as token errors. 
%Punctuation and cases that are missing, incorrect, or added in the constituent words are all counted as token errors.

\begin{table*}[th]
    \centering
    %\begin{tabular}{|p{30mm}||p{10mm}| p{20mm}|p{20mm}|p{20mm}|p{20mm}|p{20mm}|}
    \begin{tabular}{|p{30mm}||p{13mm}|p{13mm}|| p{13mm}|p{13mm}|p{13mm}|p{17mm}|p{17mm}|}
    \hline
        Dataset & Utts & Tokens & Baseline (137M) & GPT2\_XL (1.7B) & ZPP (350M) & Whisper\_L V2 (1.55B) & Whisper\_M EN (769M) \\ \hline     \hline
        commonvoice\_clean & 339 & 3352 & 8.35 & 7.73 & 8.53\ & \textbf{6.29}\ & 7.58\  \\ \hline
        commonvoice\_other & 337 & 3327 & 22.36 & 19.93 & 21.34\ & \textbf{18.33}\ & 20.47\  \\ \hline
        gigaspeech\_clean & 257 & 6049 & 11.04 & 11.19 & \textbf{10.53}\ & 11.18\ & 12.27\  \\ \hline
        gigaspeech\_other & 296 & 6227 & 24.7 & \textbf{23.40} & 24.13\ & 27.14\ & 27.08\  \\ \hline
        librispeech\_clean & 290 & 5956 & \textbf{7.94} & 8.43 & 8.34\ & 8.71\ & 8.51\  \\ \hline
        librispeech\_other & 337 & 5799 & 13.99 & 13.99 & 13.94\ & \textbf{13.09}\ & 13.23\  \\ \hline
        SPGISpeech & 1000 & 24489& 8.62 & 8.90 & \textbf{8.33}\ & 10.22\ & 9.68\  \\ \hline
        tedlium\_clean & 169 & 5869& 10.33 & \textbf{9.00} & 9.67\ & 10.14\ & 10.56\  \\ \hline
        tedlium\_other & 171 & 5665& 13.08 & 12.32 & \textbf{12.13}\ & 14.17\ & 14.16\  \\ \hline
        voxpopuli\_clean & 207 & 5042 &8.79 & 8.53 & \textbf{8.23}\ & 8.61\ & 8.43\  \\ \hline
        voxpopuli\_other & 210 & 4971 & 13.82 & 14.77 & \textbf{13.69}\ & 14.16\ & 14.12\  \\ \hline \hline
        Average & & &13.00 & \textbf{12.56} & 12.62\ & 12.91\ & 13.28\ \\ \hline
    \end{tabular}
    \caption{TER(\%) on testing sets.  }
    \label{TER}
\end{table*}

\subsection{Experimental Setup}
\subsubsection{Baseline}

The baseline model is a typical attention-based encoder-decoder (AED) model. The speech encoder has two convolutional layers that reduce the time frame by a factor of 4, followed by 24 conformer layers. Each conformer layer contains an 8-head attention mechanism and a depthwise convolution with a kernel size of 3, surrounded by two 1024-dimension feedforward layers. The decoder comprises 6 transformer layers, each with a 2048-dimension feed forward layer. The embedding dimension for both the encoder and decoder is set to 512. A SentencePiece tokenizer with 6,000 tokens is utilized to convert transcriptions into token sequences. During training, CTC and attention CE losses are employed with combination weights of 0.2 and 0.8, respectively.

\subsubsection{Adapting encoder-decoder based LM}

The schematic diagram of the encoder-decoder-based LM adaptation is shown in the Figure \ref{fig:encoder-decoder}. We use the pretrained Z-Code++ \cite{he2023zcode} base model (ZPP) as the seed for the text encoder and decoder. It was trained on 160GB of English text data using the DeBERTa \cite{he2021deberta} V2 vocabulary, which contains 128,000 tokens. To avoid setting different learning rates for the speech and text branches during training, the speech encoder parameters are also initialized using the earlier checkpoint of the baseline model. CTC and CE losses are optimized on 75,000 hours of paired speech-text data, while MLM loss is trained on the same 160GB of English text data used in Z-Code++ pretraining. We trained the entire model using CTC, CE, and MLM losses for 200,000 steps on the training data.
%However, the text data has not been fully utilized even for a single sweep.

\subsubsection{Adapting decoder-only based LM}
We utilize the GPT-2 \cite{radford2019language} as the decoder-based language model component in our system. We have experimented with two versions of GPT-2: GPT2 medium, which has 355 million parameters, and GPT2 XL, which has 1.5 billion parameters. We maintain the same speech encoder as the baseline. Regarding to the LoRA setup, we use the dimension of 8 and the LoRA alpha value of 32. To ensure effective CTC down-sampling during the initial stages of training, we pre-train the encoder using CTC loss for a brief period. Then we train the entire model using CTC and CE losses for  200,000 steps on the training data.

\subsection{Experimental Results and Analysis}

\begin{figure*}
\centering
\includegraphics[width=0.9\textwidth]{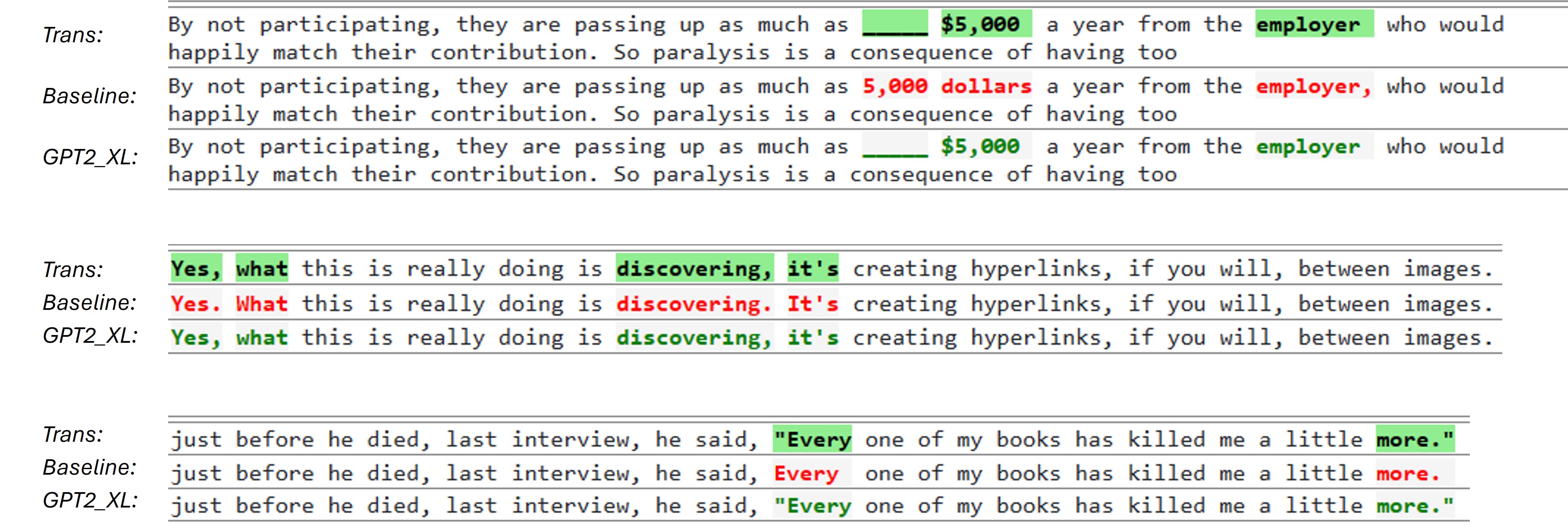}
\caption{The examples of improved utterances generated by GPT2\_XL over baseline model.}
\label{fig:example}
\end{figure*}

We compare the performance of five models in terms of TER: Baseline, GPT2\_XL (adapting a decoder-only LM), ZPP (adapting an encoder-decoder LM), Whisper\_L V2 (Whisper multi-lingual Large V2 model), and Whisper\_M En (Whisper English Medium model). The breakdown of TERs on different datasets and the average TERs over the number of datasets are shown in Table \ref{TER}, where the model size with regard to the number of parameters is indicated under the model name. Compared to the baseline model, GPT2\_XL and ZPP achieve better performance on 7 and 9 out of 11 datasets, respectively, improving the average TER from 13.00\% to 12.56\% and 12.62\%. Although Whisper\_L V2 is a multi-lingual model, its performance on our English evaluation set is better than that of Whisper\_M\_En. Both GPT2\_XL and ZPP outperform Whisper\_L V2 in terms of average TER, but Whisper\_L V2 still achieves the best TER on 3 datasets among all models.

We divide recognition errors, including substitution, deletion, and insertion, into four categories: punctuation error, capitalization error, ITN error, and lexical error. An error is defined as a punctuation or capitalization error if the predicted token and that in transcription differ only in punctuation or capitalization, respectively. ITN errors are caused by ITN, for example, ``40\%'' vs ``forty percent''. All other errors are categorized as lexical errors. We analyze the results generated by the baseline model and those of adapted LMs and observe that adapted LMs can reduce errors in all four categories. We show the error patterns of the baseline and GPT2\_XL models on the evaluation set of tedlium\_clean in Table \ref{Analysis}. Similar patterns are observed in other improved evaluation sets. We also find that the adapted LMs tend to improve the prediction performance on complex punctuation such as quotation marks. Three examples of improved utterances over the baseline model are demonstrated in Figure \ref{fig:example}.

\begin{table}[!ht]
\centering
\begin{tabular}{|l||c|c|c|c|}
\hline 
 & Punc & Cap & ITN & Lexical \\
\hline \hline
Baseline & 284 & 115 & 3 & 204 \\
\hline
GPT2\_XL & 260 & 110 & 3 & 155 \\
\hline
\end{tabular}
\caption{The number of punctuation, capitalization, ITN and lexical errors produced by Baseline and GPT2\_XL models on the tedlium\_clean evaluation set.}
\label{Analysis}
\end{table}

\begin{table}[!ht]
    \centering
    \begin{tabular}{|l||l|l|l|}
    \hline
        Dataset & Baseline & GPT2\_M & zpp\_token   \\ 
         & 137M & 460M & 325M  \\ 
        \hline \hline
        common\_voice\_clean & 8.35 & 8.74 & 9.61  \\ \hline
        common\_voice\_other  & 22.36 & 21.28 & 23.35  \\ \hline
        gigaspeech\_clean & 11.04 & 11.39 & 11.04  \\ \hline
        gigaspeech\_other & 24.7 & 24.51 & 24.84  \\ \hline
        librispeech\_clean & 7.94 & 8.51 & 8.41  \\ \hline
        librispeech\_other & 13.99 & 13.80 & 15.11  \\ \hline
        SPGISpeech & 8.62 & 8.93 & 9.22  \\ \hline
        tedlium\_clean & 10.33 & 9.18 & 9.80  \\ \hline
        tedlium\_other & 13.08 & 12.66 & 12.94  \\ \hline
        voxpopuli\_clean & 8.79 & 8.41 & 8.59  \\ \hline
        voxpopuli\_other & 13.82 & 14.26 & 13.96  \\ \hline \hline
        Average & 13.00 & 12.88 & 13.35  \\ \hline
    \end{tabular}
\caption{Ablation study}
\label{ablation}
\end{table}

\subsection{Ablations}

We conducted an ablation study and presented the results in the table \ref{ablation}. First, we investigated the impact of LLM size on speech recognition performance. Although the study is only carried out on decoder only based LM, we believe the conclusion could also generalize to encoder-decoder based LM. Our results show that GPT2 medium (GPT2\_M), despite only containing 27\% of the parameters of GPT2\_XL, can still improve performance over the baseline model. But with increasing in LLM size, as exemplified by the GPT2\_XL in table \ref{TER}, the result can become better. This suggests that exploring larger LLMs could be a promising direction for future research.

%LORA can significantly reduce the number of trainable parameters compared to full model finetuning. We hypothesize that the speech encoder may be adapted to generate embeddings closer to the token embeddings used in GPT, rather than adapting GPT to take speech inputs directly, since we train speech encoder and LORA adapter to GPT in an end-to-end manner. 

We then studied whether leveraging only the tokenizer from the LM could enhance performance. The column labeled "zpp\_token" is the system by simply replacing baseline tokenizer with zpp tokenizer, and the results show that while the model size increased significantly with the Z-Code++ tokenizer, the performance in terms of TER actually degraded. This indicates that LLM tokenizer is not the key to improving for our proposed system, and only leveraging the tokenizer is not helpful.

%We think this may be due to insufficient training of such a large tokenizer with 75000 hours of data.

\section{Conclusion}
In this paper, we exploit the capabilities of pretrained LLMs by adapting them with speech for fully formatted end-to-end ASR transcription. We study two methods, applied to either encoder-decoder-based LMs or decoder-only-based LMs, to improve ASR models trained using large-scale supervised data. Our experimental results show that this is a promising research direction. In the future, we plan to explore even larger-scale LLMs to further enhance the performance of speech processing tasks.

\bibliographystyle{IEEEbib}
\bibliography{refs}

\end{document}